\documentclass[twocolumn]{aastex62}

\newcommand{\Mh}{M_{\rm h}}
\newcommand\Msun{M_{\odot}}
\newcommand\Mmin{M_{\min}}
\newcommand\Mstar{M_{\star}}

\newcommand\zphot{z_{\rm phot}}
\newcommand\zspec{z_{\rm spec}}
\newcommand\fsat{f_{\rm s}}
\newcommand\bg{b_{\rm g}}
\newcommand\bh{b_{\rm h}}
\newcommand\sigmaM{\sigma_{\log{M}}}
\newcommand\logMstarlimit{\log_{10}(M_{\star, {\rm limit}}/h^{-2}\Msun)}

\newcommand\logMstar{\log_{10}(M_{\star}/\Msun)}
\newcommand\Mstarlimit{M_{\star, {\rm limit}}}

\shorttitle{Halo-model analysis of the HSC LRGs}
\shortauthors{Ishikawa et al.}

\begin{document}

\title{Halo-model analysis of the clustering of photometric luminous red galaxies at $0.10 \leq z \leq 1.05$ from the Subaru Hyper Suprime-Cam Survey}

\correspondingauthor{Shogo Ishikawa}
\email{shogo.ishikawa.astro@gmail.com}

\author{Shogo Ishikawa}
\affil{Center for Computational Astrophysics, National Astronomical Observatory of Japan, Mitaka, Tokyo 181-8588, Japan}
\affil{National Astronomical Observatory of Japan, Mitaka, Tokyo 181-8588, Japan}
\author{Teppei Okumura}
\affil{Institute of Astronomy and Astrophysics, Academia Sinica, No.~1, Section~4, Roosevelt Road, Taipei 10617, Taiwan}
\affil{Kavli Institute for the Physics and Mathematics of the Universe (WPI), UTIAS, The University of Tokyo, Kashiwa, Chiba 277-8583, Japan}
\author{Masamune Oguri}
\affil{Kavli Institute for the Physics and Mathematics of the Universe (WPI), UTIAS, The University of Tokyo, Kashiwa, Chiba 277-8583, Japan}
\affil{Research Center for the Early Universe, The University of Tokyo, 7-3-1 Hongo, Bunkyo-ku, Tokyo, 113-0033, Japan}
\affil{Department of Physics, The University of Tokyo, 7-3-1 Hongo, Bunkyo-ku, Tokyo 113-0033, Japan}
\author{Sheng-Chieh Lin}
\affil{Institute of Astronomy and Astrophysics, Academia Sinica, No.~1, Section~4, Roosevelt Road, Taipei 10617, Taiwan}
\affil{Department of Physics and Astronomy, University of Kentucky, 505 Rose Street, Lexington, KY 40506, USA.}

\begin{abstract}
We present the clustering analysis of photometric luminous red galaxies (LRGs) at a redshift range of $0.1\leq z\leq1.05$ using $615,317$ photometric LRGs selected from the Hyper Suprime-Cam Subaru Strategic Program covering $\sim124$ deg$^{2}$. 
Our sample covers a broad range of stellar masses and photometric redshifts and enables a halo occupation distribution analysis to study the redshift and stellar-mass dependence of dark halo properties of LRGs. 
We find a tight correlation between the characteristic dark halo mass to host central LRGs, $\Mmin$, and the number density of LRGs independently of redshifts, indicating that the formation of LRGs is associated with the global environment. 
The $\Mmin$ of LRGs depends only weakly on the stellar mass $\Mstar$ at $\Mstar\lesssim10^{10.75}h^{-2}\Msun$ at $0.3<z<1.05$, in contrast to the case for all photometrically selected galaxies for which $\Mmin$ shows significant dependence on $\Mstar$ even at low $\Mstar$. 
The weak stellar mass dependence is indicative of the dark halo mass being the key parameter for the formation of LRGs rather than the stellar mass. 
Our result suggests that the halo mass of $\sim10^{12.5\pm0.2}h^{-1}\Msun$ is the critical mass for an efficient halo quenching due to the halo environment. 
We compare our result with the result of the hydrodynamical simulation to find that low-mass LRGs at $z\sim1$ will increase their stellar masses by an order magnitude from $z=1$ to $0$ through mergers and satellite accretions, and a large fraction of massive LRGs at $z<0.9$ consist of LRGs that are recently migrated from massive green valley galaxies or those evolved from less massive LRGs through mergers and satellite accretions. 
\end{abstract} 

\keywords{cosmology: observations --- dark matter --- large-scale structure of universe --- galaxies: evolution --- formation}

\section{Introduction} \label{sec:intro}
The formation of the large-scale structure of the Universe, which can be traced by galaxies, is largely governed by cosmology in the early Universe. 
Small matter density fluctuations in the early Universe evolve into biased objects, known as dark halos, via gravitational instability. 
Galaxies, which are an important tracer of the large-scale structure, form in dark halos by trapping baryons by their gravitational wells. 
Dark halos formed in the early epoch induce cooling and condense processes on the accreted baryonic gasses at the early stage of the Universe. Therefore, highly biased galaxies tend to reside in old, massive dark halos \citep[e.g.,][]{white78,blumenthal84}. 

Luminous red galaxies (LRGs) are thought to be a passively evolving, long-lived galaxy population hosted by such old dark halos \citep{eisenstein01}. 
Given the large bias and their brightness, LRGs are a useful tracer of the large-scale structure of the Universe. 
Many redshift surveys have constructed large samples of LRGs over a broad redshift range, especially after the notable success by the Sloan Digital Sky Survey (SDSS), from which the clustering of LRGs has been measured as a function of various baryonic properties of the LRGs \citep[e.g.,][]{zheng09,reid10,zehavi11,guo13}. 

Since all galaxies are considered to form within halos in the current paradigm of cosmic structure formation, investigating galaxy clustering provides clues to understand the relationship between galaxies and host dark halos and reveal properties of the underlying dark matter distribution. 
Furthermore, measuring and interpreting the galaxy clustering signals play an important role in discriminating different physical processes to drive galaxy formation and evolution \citep[e.g.,][]{wechsler18}. 

Two-point auto-correlation functions (2PCFs) are commonly used to quantify the clustering of galaxies \citep{totsuji69,peebles80}. 
The 2PCF provides various information at different physical scales ranging from the large-scale structure of the Universe to the galaxy formation within dark halos. 
At large physical scales, the 2PCF is approximately attributed to the matter-matter correlation function in the linear regime enhanced by the halo bias \citep[e.g.,][]{seljak00,ma00}. 
On the other hand, the 2PCF at small scales contains a wealth of information on the non-linear physical processes of galaxy formation and evolution, and the central--satellite interactions within dark halos \citep[e.g.,][]{kravtsov04,tinker05}. 

Observed 2PCFs of galaxies can be interpreted using analytical halo models that are developed based on the galaxy formation model in the context of the $\Lambda$-dominated Cold Dark Matter ($\Lambda$CDM) cosmological model \citep[][for a review]{cooray02}. 
One of the most successful halo models for analyzing observed clustering signals is the halo occupation distribution (HOD) model \citep[e.g.,][]{berlind02,berlind03,vdbosch03}. 
The HOD model characterizes the galaxy bias in terms of the galaxy occupation within dark halos, and predicts 2PCFs via the conditional probability of the number of galaxies $N$ in each halo as a function of the dark halo mass $\Mh$, $P(N|\Mh)$. 
The HOD model has successfully been applied to LRGs selected by several spectroscopic galaxy redshift surveys to reveal a biased relation between LRGs and the underlying dark matter as well as the redshift evolution of the LRG itself \citep[e.g.,][]{zheng09,white11,guo13,parejko13,zhai17}. 
However, these galaxy redshift surveys focus only on luminous galaxies and hence they can cover only the limited redshift and stellar-mass ranges.

In this paper, we report clustering properties of photometric LRGs selected by the Subaru Telescope Hyper Suprime-Cam Subaru Strategic Program \citep[HSC SSP; ][]{aihara18}. 
HSC SSP is a deep and wide optical imaging survey utilizing the capability of the wide-field imaging camera Hyper Suprime-Cam \citep[HSC; ][]{miyazaki18} mounted on the Subaru Telescope. 
Deep and wide-field photometric data of the HSC SSP and the sophisticated LRG-selection algorithm \citep[CAMIRA; ][]{oguri14} enable us to construct an LRG sample for a wide range of redshift and stellar mass, where redshifts are estimated using the photometric redshift (photo-$z$) technique. 
The main purpose of this paper is to study properties of LRGs as a function of the stellar mass and redshift through the clustering and the subsequent HOD analyses, and compare the result with that for photo-$z$-selected all galaxy samples containing both red and blue galaxies, where the latter result is also obtained in the same HSC SSP dataset \citep{ishikawa19}.

This paper is organized as follows. 
In Section~\ref{sec:data}, we present the details of our photometric data and LRG selection method using the CAMIRA algorithm \citep{oguri14,oguri18a}. The clustering and HOD analyses of the LRG samples are shown in Section~\ref{sec:clustering}. 
Results obtained by the HOD analyses of the clustering of LRGs and comparison with the all galaxy samples are given in Section~\ref{sec:result}, discussion based on the HOD analysis is presented in Section~\ref{sec:discussion}. We give a conclusion in Section~\ref{sec:summary}. 

We employ the Planck 2015 cosmological parameters \citep{planck15}; i.e., the matter, baryon, and dark energy density parameters are $\Omega_{\rm m}=0.309$, $\Omega_{\rm b}=0.049$, and $\Omega_{\rm \Lambda}=0.691$, respectively, the dimensionless Hubble parameter is $h=0.677$, the amplitude of the linear power spectrum averaged over $8h^{-1}$ Mpc scale is $\sigma_{8} = 0.816$, and the scalar spectrum index of the primordial power spectral is $n_{\rm s} = 0.967$. 
Throughout this paper, dark halo masses and stellar masses are denoted as $\Mh$ and $\Mstar$ with their units of $h^{-1}\Msun$ and $h^{-2}\Msun$, respectively. 
All of logarithm in this paper are common logarithm with base $10$. 

\section{Data and Sample Selection} \label{sec:data}
An LRG catalog used in this study is obtained from the photometric data of the HSC SSP S16A Wide layer, which covers $\sim 174$ deg$^{2}$ in total \citep{aihara18}. 
LRGs are selected by \citet{oguri18a,oguri18b} using the {\sc CAMIRA} algorithm \citep{oguri14}. 
The CAMIRA fits magnitudes and colors of galaxies with a stellar population synthesis (SPS) model of \citet{bc03} with a fixed formation redshift of $z_{\rm f}=3$ and a prior on the metallicity that depends on the stellar mass. 
The SPS model is designed to reproduce red-sequence in clusters of galaxies, and is also carefully calibrated such that it reproduce colors of spectroscopic LRGs accurately. 
\citet{oguri18b} apply this method to construct a photometric LRG catalog in HSC SSP S16A by selecting galaxies that are fitted well by the SPS model for LRGs. 
Additionally, in this paper we impose the {\tt bright-star mask} flag to avoid noise signals originated mainly from saturated and crosstalk pixels around luminous stars on clustering signals. 
Furthermore, we manually mask noisy regions and edge of the survey fields by visual to obtain a reliable LRG catalog for clustering measurements. 
The final area of our survey field is $\sim 124.31$ deg$^{2}$ and it contains $615,317$ LRGs at $0.1 \leq z \leq 1.05$. 

In the CAMIRA LRG catalog, the stellar mass and photo-$z$ of each LRG, which are derived by the CAMIRA algorithm \citep{oguri14}, are available. 
The accuracy of photo-$z$'s of CAMIRA LRGs is tested in \citet{oguri18b} by comparing them with the spectroscopic redshifts that have already been obtained by other surveys. 
The scatter of photo-$z$'s, which is defined as a scatter of $(z_{\rm phot} - z_{\rm spec})/(1 + z_{\rm spec})$ after $3\sigma$ clipping is found to be $\sim 2\%$, which is comparable to the accuracy of photo-$z$'s of LRGs selected by the redMaGiC algorithm \citep{rozo16}. 
At $0.1<z_{\rm phot}<1.05$, the outlier rate of the whole LRG samples is found to be $\sim 7\%$. 
Interested readers are referred to \citet{oguri14} and \citet{oguri18a,oguri18b} for more details of the original {\sc CAMIRA} LRG catalog and its photometric-redshift performance. 

\begin{figure}[tbp]
\epsscale{1.0}
\plotone{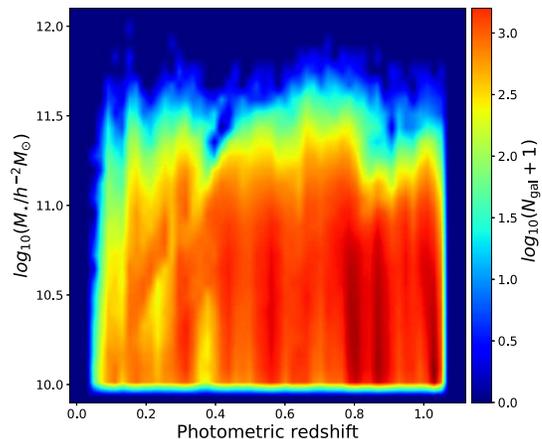}
\caption{The stellar mass versus redshift diagram of the HSC SSP S16A photometric LRG sample used in this paper. Color indicates the differential abundance of LRGs in logarithmic scale at each cell. }
\label{fig:Mstar-z}
\end{figure}

The LRG sample is divided into subsamples according to their stellar masses and photo-$z$'s. 
The stellar mass versus redshift diagram is shown in Figure~\ref{fig:Mstar-z}. 
We adopt the redshift binning similar to that of \citet{ishikawa19} for easier comparisons of physical characteristics of our LRGs with those of photo-$z$-selected all galaxy samples containing both red and blue galaxies (hereafter photo-$z$ galaxies) defined in \citet{ishikawa19} from the same dataset of HSC SSP S16A Wide layer as used in this paper. 
The redshift distribution of LRGs is shown in the top panel of Figure~\ref{fig:Nz} and details of each subsample are presented in Table~\ref{tab:nz}. 

In addition, we also generate a random-point catalog that covers the entire field of the LRG catalog. 
The surface number density of the random catalog is set to $\sim 35$ arcmin$^{-2}$, which is $\sim 100$ times larger than that of the LRG catalog, in order to reduce the Poisson noise on clustering signals. 
The random points around bright stars and near edges of the fields are excluded in the same manner as in the LRG catalog. 

\begin{table*}[tbp]
\begin{center}
\caption{Details of Cumulative Stellar-mass Limited Subsamples}
\begin{tabular}{lccccccccccccccc} \hline \hline
& \multicolumn{3}{c}{$z_{1}$} & & \multicolumn{3}{c}{$z_{2}$} & & \multicolumn{3}{c}{$z_{3}$} & & \multicolumn{3}{c}{$z_{4}$} \\ 
& \multicolumn{3}{c}{$0.10 \leq z < 0.30$} & & \multicolumn{3}{c}{$0.30 \leq z < 0.55$} & & \multicolumn{3}{c}{$0.55 \leq z < 0.80$} & & \multicolumn{3}{c}{$0.80 \leq z \leq 1.05$}\\ 
\cline{2-4}\cline{6-8}\cline{10-12}\cline{14-16}
Stellar-mass limit\footnote{Threshold stellar mass of each subsample in units of $h^{-2}\Msun$ in a logarithmic scale.} & \multicolumn{1}{c}{$N$\footnote{The number of LRGs of each subsample. }} & \multicolumn{1}{c}{$M_{\star, {\rm med}}$\footnote{Median stellar mass of each subsample in units of $h^{-2}\Msun$ in a logarithmic scale. }} & \multicolumn{1}{c}{$n_{g}$\footnote{LRG number density in units of $10^{-3} h^{-3}$Mpc$^{3}$ }} & & \multicolumn{1}{c}{$N$} & \multicolumn{1}{c}{$M_{\star, {\rm med}}$} & \multicolumn{1}{c}{$n_{g}$} & & \multicolumn{1}{c}{$N$} & \multicolumn{1}{c}{$M_{\star, {\rm med}}$} & \multicolumn{1}{c}{$n_{g}$} & & \multicolumn{1}{c}{$N$} & \multicolumn{1}{c}{$M_{\star, {\rm med}}$} & \multicolumn{1}{c}{$n_{g}$} \\ \hline
$10.00$ & $56,249$ & $10.53$ & $9.12$ & & $132,223$ & $10.59$ & $5.07$ & & $196,623$ & $10.57$ & $3.76$ & & $230,222$ & $10.48$ & $3.07$ \\
$10.25$ & $41,586$ & $10.67$ & $6.50$ & & $107,664$ & $10.68$ & $4.10$ & & $158,538$ & $10.67$ & $3.03$ & & $170,292$ & $10.61$ & $2.24$ \\
$10.50$ & $29,693$ & $10.79$ & $4.55$ & & $78,205$ & $10.80$ & $3.17$ & & $112,600$ & $10.80$ & $2.16$ & & $110,352$ & $10.74$ & $1.45$ \\
$10.75$ & $16,863$ & $10.96$ & $2.48$ & & $45,071$ & $10.95$ & $1.67$ & & $65,009$ & $10.96$ & $1.22$ & & $52,779$ & $10.90$ & $0.66$ \\
$11.00$ & $6,967$ & $11.13$ & $1.00$ & & $17,740$ & $11.12$ & $0.64$ & & $27,035$ & $11.14$ & $0.51$ & & $13,726$ & $11.09$ & $0.17$ \\ \hline\hline
\label{tab:nz}
\end{tabular}
\end{center}
\end{table*}

\section{Clustering and HOD Analysis} \label{sec:clustering}
\subsection{Angular Auto-correlation Function} \label{subsec:acf}
We measure angular auto-correlation functions (ACFs) of the LRG samples. 
ACFs are calculated using an estimator proposed by \citet{landy93} as:
\begin{equation}
\omega\left(\theta\right) = \frac{{\rm DD} - 2 {\rm DR} + {\rm RR}}{{\rm RR}}, 
\label{eq:acf}
\end{equation}
where ${\rm DD}$, ${\rm DR}$, and ${\rm RR}$ denote the numbers of normalized pairs of galaxy--galaxy, galaxy--random, and random--random within the separation angle range of $\theta \pm \delta\theta$, respectively. 
In this paper, we measure ACFs in the angular scale range of $-3.4 \leq \log_{10}(\theta) \leq 0.0$ with the separation of $\log_{10}(\delta\theta)=0.1$ in a degree scale. 

An observed galaxy correlation function is underestimated at large-angular scales due to the finite survey field, known as the integral constraint (IC) \citep[e.g.,][]{groth77}. 
We first calculate this effect by Monte Carlo integration as: 
\begin{equation}
{\rm IC} = \frac{\sum_{i}\theta^{1-\gamma}_{i}{\rm RR\left(\theta_{i}\right)}}{\sum_{i} {\rm RR\left(\theta_{i}\right)}}, 
\label{eq:ic}
\end{equation}
assuming that the ACF can be modeled by the power-law form $\omega(\theta)\propto \theta^{-\gamma}$ \citep{roche99}. 
The unbiased ACFs, $\omega_{{\rm true}}$, can be evaluated by correcting the observed ACFs, $\omega_{{\rm obs}}$, as:
\begin{equation}
\omega_{{\rm true}} \left(\theta \right) = \omega_{{\rm obs}} \left(\theta \right) \frac{\theta^{1-\gamma}}{\theta^{1-\gamma} - {\rm IC}}.
\label{eq:w_true}
\end{equation}

We use the jackknife resampling method to evaluate errors of the ACFs \citep[e.g.,][]{norberg09}. 
We divide our survey field into $123$ subfields, each of which covers $\sim1$ deg$^{2}$, and calculate the ACFs $123$ times, removing each subfield. 
The covariance matrix can be computed as: 
\begin{equation}
C_{ij} = \frac{N-1}{N} \sum_{k=1}^{N} \left( \omega_{k}\left(\theta_{i}\right) - \bar{\omega} \left(\theta_{i}\right) \right) \left( \omega_{k}\left(\theta_{j}\right) - \bar{\omega} \left(\theta_{j}\right) \right), 
\label{eq:cm}
\end{equation}
where $N=123$, $C_{ij}$ is an $(i, j)$ element of the covariance matrix, $\omega_{k}\left(\theta_{i}\right)$ is the ACF of the $i$th angular bin of the $k$th jackknife realization, and $\bar{\omega} \left(\theta_{i}\right)$ is 
the ACF with the $i$th angular bin averaged over the $N$ realizations, $\bar{\omega} \left(\theta_{i}\right)=N^{-1}\sum_{k=1}^N\omega_k\left(\theta_{i}\right)$. 

\subsection{HOD Analysis} \label{subsec:hod}
\subsubsection{Methodology} \label{subsubsec:hod_method}
We use an HOD formalism for interpreting the observed ACFs and link the LRGs to their host dark halos \citep[e.g.,][]{seljak00,berlind02}. 
The HOD model parameterizes the occupation of galaxies as a function of dark halo mass and predicts galaxy correlation functions according to the assumed galaxy distribution within dark halos. 
In this study, we adopt the standard galaxy occupation function proposed by \citet{zheng05} for the distribution of our LRG samples within dark halos. 
The total number of LRGs within a dark halo with mass $\Mh$, $N_{\rm tot}(\Mh)$, can be decomposed into the central and satellite components, $N_{\rm c}$ and $N_{\rm s}$, respectively, and 
is described as: 
\begin{equation}
N_{\rm tot}\left(\Mh \right) = N_{\rm c}\left(\Mh\right) \left[ 1 + N_{\rm s}\left(\Mh\right) \right]. 
\label{eq:hof}
\end{equation}
In the HOD model of \citet{zheng05}, the occupations of central and satellite galaxies are written as:
\begin{eqnarray}
N_{\rm c}\left(\Mh\right) &=& \frac{1}{2} \left[ 1 + {\rm erf}\left(\frac{\log_{10}{\Mh} - \log_{10}{\Mmin}}{\sigmaM}\right) \right],
\label{eq:Nc} \\
N_{\rm s}\left(\Mh\right) &=& \left(\frac{\Mh - M_{0}}{M_{1}}\right)^{\alpha}. 
\label{eq:Ns}
\end{eqnarray}
There are five free HOD parameters in the above occupation functions; $\Mmin$ is the characteristic mass to host a central galaxy, $M_{1}$ is a mass for a halo with a central galaxy to host one satellite, $M_{0}$ is the mass scale to truncate satellites, $\sigmaM$ is the characteristic transition width, and $\alpha$ is the slope of the power law for the satellite HOD. 
Once a set of these parameters is given, one can uniquely compute the three-dimensional power spectrum, which is then converted to the angular correlation function, $\omega_{\rm HOD}(\theta)$. 
Previous studies successfully reproduced galaxy clustering of LRGs \citep[e.g.,][]{zheng09,zhai17} and massive red galaxies \citep[e.g.,][]{brown08,matsuoka11} at $z<1$ using the above occupation functions. 

The HOD parameters are constrained by comparing the observed ACF with the predicted one from the HOD model using the $\chi^{2}$ statistic. 
The $\chi^{2}$ is computed as: 
\begin{eqnarray}
\chi^{2} & = & \sum_{i, j} \left[ \omega_{\rm true}\left(\theta_{i}\right) - \omega_{\rm HOD}\left(\theta_{i}\right) \right] \left( C_{ij}^{-1} \right) \left[ \omega_{\rm true}\left(\theta_{j}\right) - \omega_{\rm HOD}\left(\theta_{j}\right) \right] \nonumber \\
& + & \frac{\left( n_{g}^{\rm obs} - n_{g}^{\rm HOD} \right)^{2}}{{\sigma_{n_{g}}}^{2}}, 
\label{eq:chi_hod}
\end{eqnarray}
where $\omega_{\rm HOD}\left(\theta_{i}\right)$ is the ACFs of $i$th angular bin from the HOD model, $n_{g}^{\rm obs}$ is the observed number density of galaxies (see Table~\ref{tab:nz}), $\sigma_{n_{g}}$ is its uncertainty, and $n_{g}^{\rm HOD}$ is the number density predicted by the HOD model
\begin{equation}
n_{g}^{\rm HOD} = \int d\Mh \frac{dn}{d\Mh} N\left( \Mh \right), 
\label{eq:ng_hod}
\end{equation}
where $dn/d\Mh$ is a halo mass function. 
In calculating the inverse covariance matrix, $C_{ij}^{-1}$, from equation~(\ref{eq:cm}), we apply a correction factor presented by \citet{hartlap07} to avoid underestimating the inverse covariance due to the finite realization effect. 

The uncertainty of the observed number density, $\sigma_{n_{g}}$, takes account of the effect of the photo-$z$ errors. 
While it has been confirmed that the CAMIRA LRG samples are less affected by photo-$z$ uncertainties compared to all photo-$z$ galaxies, we conservatively introduce $10\%$ uncertainties on the galaxy abundance that takes account of photo-$z$ errors and some of unknown systematic biases as is the case with \citet{zhou20}. 

To predict ACFs from the HOD framework, one needs to calculate several quantities analytically. 
We employ a halo mass function proposed by \citet{sheth99}, an NFW profile \citep{navarro97} as a density profile of dark halos, the mass and redshift dependence of a concentration parameter presented by \citet{takada03}, and a large-scale halo bias of \citet{tinker10} with a halo exclusion effect \citep{zheng04,tinker05}. 
We use the non-linear power spectrum of \citet{smith03} with a matter transfer function of \citet{eisenstein98}. 

\subsubsection{Photo-$z$ error estimation} \label{subsubsec:nz}
\begin{figure}[tbp]
\plotone{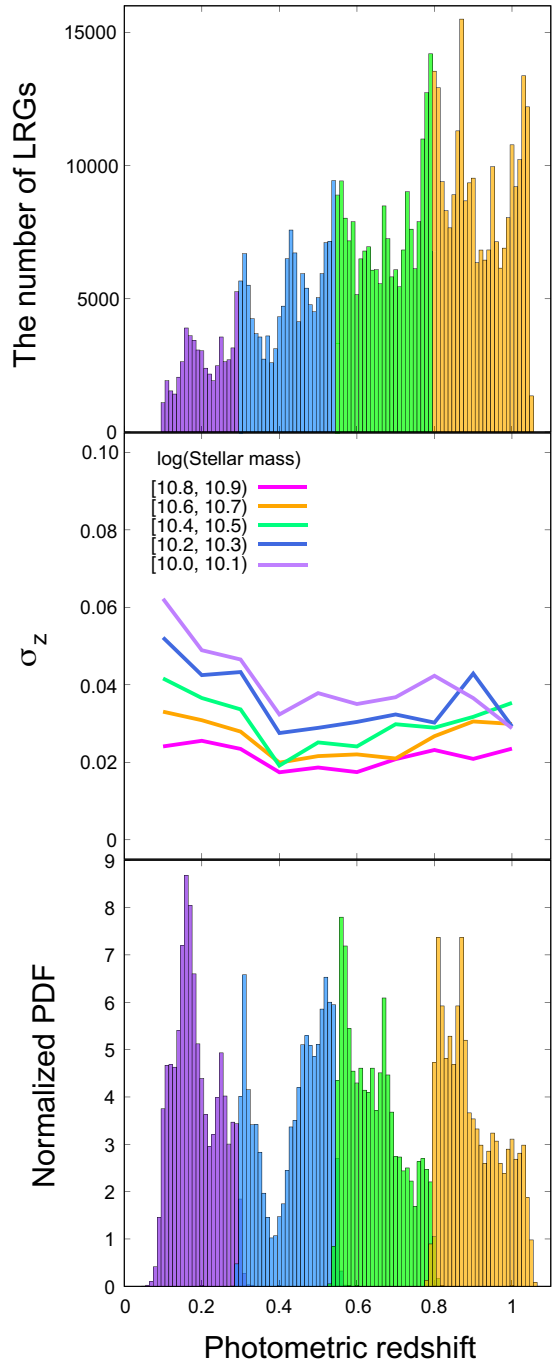}
\caption{Top: Redshift distribution of the LRG sample. 
LRGs are divided into four redshift bins: $0.10 \leq z < 0.30$ (purple), $0.30 \leq z < 0.55$ (blue), $0.55 \leq z < 0.80$ (green), and $0.80 \leq z \leq 1.05$ (orange). 
Middle: Scatter of photo-$z$'s as a function of photo-$z$. 
These scatters are evaluated after excluding outliers (see text for more details). 
Different colors indicate results for different stellar-mass slices. Bottom: Redshift distribution of total LRGs at each redshift bin that takes account of the photo-$z$ errors. 
Colors indicate the redshift bins defined in the top panel. }
\label{fig:Nz}
\end{figure}

Although the redshift distributions of LRGs using the best-fitting photo-$z$'s are already evaluated in the top panel of Figure~\ref{fig:Nz}, they are likely different from the true redshift distributions of LRGs due to photo-$z$ uncertainties. 
Therefore, it is essential to use redshift distributions that take full account of photo-$z$ errors in the HOD-model analysis to fit observed ACFs. 
To take account of photo-$z$ errors that induce tails of the distribution at each redshift bin boundary in the interpretation of observed clustering, we recalculate redshift distributions of LRGs by considering those uncertainties. 

\begin{figure*}[tbp]
\plotone{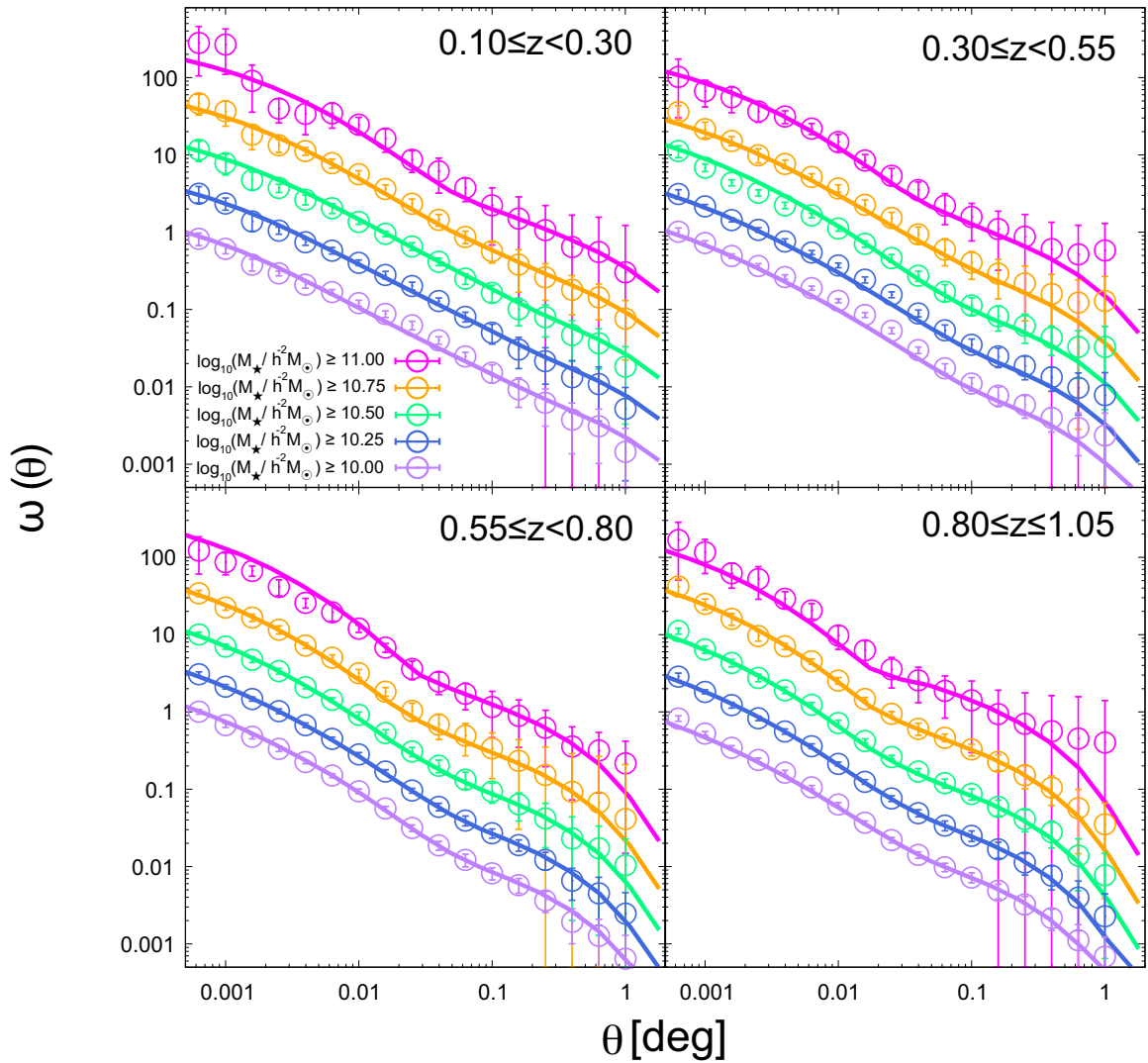}
\caption{Observed ACFs of LRGs (circles) and their best-fitting ACFs derived by the HOD model (solid lines) at $0.10 \leq z < 0.30$ (top left panel), $0.30 \leq z < 0.55$ (top right panel), $0.55 \leq z < 0.80$ (bottom left panel), and $0.80 \leq z \leq 1.05$ (bottom right panel) redshift bins, respectively. In the HOD fitting procedure, correlations between angular bins are taken into consideration by covariance matrices calculated by the jackknife resampling method. Amplitudes of ACFs are shifted arbitrarily for an illustrative purpose. }
\label{fig:HOD}
\end{figure*}

The redshift distribution of the LRGs including photo-$z$ errors is evaluated as follows. 
First, we focus on LRGs whose spectroscopic redshifts have already been measured by other surveys and calculate the scatter of photometric redshifts $\sigma_{z}$ of the residual of both redshifts, i.e., $(\zphot - \zspec)/(1 + \zspec)$, where $\zphot$ and $\zspec$ respectively refer to photometric and spectroscopic redshifts of each LRG. 
It should be noted that the LRGs with spec-$z$ information account for only $\sim 5.3 \%$ of the total LRG sample. 
The photo-$z$ scatter is evaluated by calculating the root mean square of the residual after excluding outliers that satisfy $|\zphot - \zspec|/(1 + \zspec) > 0.15$. 
The fraction of outlier LRGs among those with spec-$z$'s is $\sim 2.2 \%$. 
The scatter of photo-$z$ is estimated as a function of both stellar mass and redshift bins with bin sizes of $\delta \logMstar = 0.1$ and $\delta z = 0.1$, respectively. 
The middle panel of Figure~\ref{fig:Nz} shows the scatter of photo-$z$'s at various stellar-mass slices as a function of photo-$z$. 
As expected, less massive and high-$z$ LRGs tend to have larger photo-$z$ uncertainties. 

Using the estimated photo-$z$ scatter, photo-$z$'s of individual LRGs are randomly reassigned assuming the Gaussian distribution. 
We repeat this procedure $100$ times in order to obtain averaged redshift distributions including the photo-$z$ error of each LRG. 
The bottom panel of Figure~\ref{fig:Nz} shows the LRG redshift distribution for each redshift bin obtained by the above procedure. 
These redshift distributions including the photo-$z$ errors are used in the following HOD analysis. 

\subsubsection{HOD-model fitting} \label{subsubsec:hod_fitting}
We explore the HOD parameters of each subsample that reproduce the observed ACFs. 
The HOD parameters are constrained using a population Monte Carlo \citep[PMC;][]{wraith09} algorithm. 
The PMC is an importance sampling method that obtains new samples from a proposal distribution, and the posterior updates the proposal distribution iteratively. 
We use the {\sc CosmoPMC} package \citep{kilbinger11} to derive mean values and $1\sigma$ confidence intervals of the HOD parameters. 
The confidence region is evaluated  by integrating the normalized posterior of the final iteration from the mean values to the points that reach $\pm63.27/2$\%. 
  
Results of the HOD-model fitting are shown in Figure~\ref{fig:HOD} and the best-fitting parameters are listed in Table~\ref{tab:hod_params}. 
The HOD model successfully reproduces the observed ACFs of LRGs over the whole ranges of the angular scales, the stellar mass, and the redshift explored in this paper. 
The HOD halo mass parameters $\Mmin$ and $M_{1}$ are tightly constrained thanks to the accurate photo-$z$'s as well as the large sample size achieved by the large survey volume of the HSC SSP. 

\section{Results} \label{sec:result}
In this section, we present constraints on physical parameters of LRGs and their host halos based on the HOD modeling. 
We show constraints on the halo mass parameters, $\Mmin$ and $M_1$, in Figure~\ref{fig:mh}, the satellite fraction of LRGs, $\fsat$, in Figure~\ref{fig:fsat}, and the large-scale galaxy bias, $\bg$, in Figure~\ref{fig:bg}. 
In each figure, the left panel compares our constraints to those for LRGs in the literature based on different surveys \citep{ross07,ross08,brown08,zheng09,white11,zehavi11,nikoloudakis13,parejko13,zhai17} as a function of the galaxy number density. 
It is noted that the halo occupation function of central LRGs adopted by \citet{zehavi11} and \citet{nikoloudakis13} are different from that in the other studies. 
The right panel compares our constraints to those for the photo-$z$ galaxies including both red and blue galaxies from the same HSC survey \citep{ishikawa19} as a function of the stellar mass threshold. 
We will discuss these results in more detail in the following subsections. 

\subsection{$\Mmin$ and $M_{1}$} \label{subsec:mmin-m1}
Here we investigate the redshift evolution of the HOD halo mass parameters, $\Mmin$ and $M_{1}$. 
By definition, $\Mmin$ is a halo mass for which the expected number of central galaxy occupation is $50\%$, whereas $M_{1}$ is a typical mass of a halo that is expected to possess one satellite galaxy. 
We do not discuss the other halo mass parameter in our HOD model, $M_{0}$, since it has large uncertainties ($\sim \pm 2$ dex errors in typical cases).

\begin{figure*}[htbp]
\plotone{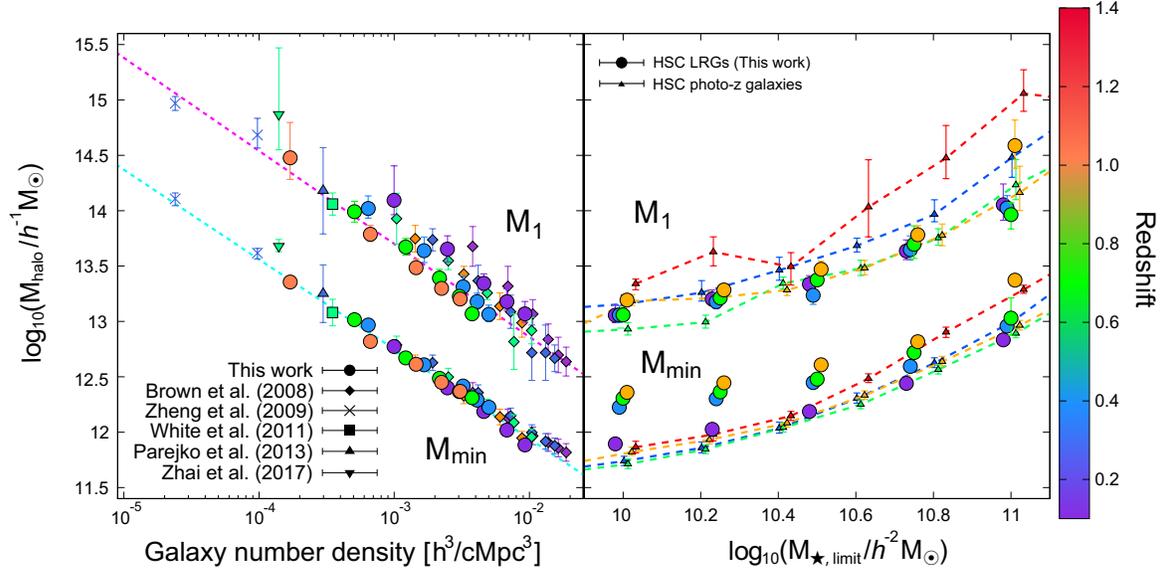}
\caption{Left: Observed HOD halo mass parameters ($\Mmin$ and $M_{1}$ in logarithmic scales) of LRGs as a function of the number density of galaxies. Different colors show the different redshift range as indicated by the color bar on the right side of the figure, both for our results and results in the literature. Dashed cyan and magenta lines are best-fitted power-law slopes for our results of $\Mmin$ and $M_{1}$, respectively. Right: HOD halo mass parameters ($\Mmin$ and $M_{1}$ in logarithmic scales) of LRGs (circles) and HSC photo-$z$ galaxies (triangles) as a function of the stellar-mass threshold. Different colors show the different redshift ranges. Symbols are slightly shifted along the horizontal axis for illustrative purpose. }
\label{fig:mh}
\end{figure*}

First, we compare the constraints on $\Mmin$ and $M_{1}$ for CAMIRA LRGs with those for LRGs from previous HOD studies to check the consistency among them. 
Thanks to the wide survey area and deep imaging data of the HSC SSP Wide layer, we successfully determine the HOD halo mass parameters for wide stellar mass and redshift ranges as shown in the left panel of Figure~\ref{fig:mh}. 
Our constraints, particularly on $\Mmin$, show excellent agreement with those of the previous studies. 
The constrained values of $\Mmin$ and $M_{1}$ are tightly correlated with the number density of LRGs and approximately follow power-law relations with negative slopes, $\Mmin \propto n_{g}^{-0.8}$ and $M_{1} \propto n_{g}^{-1.0}$, which are depicted by the cyan and magenta lines, respectively. 
The difference of their slopes indicates that the formation efficiency of massive satellite LRGs is much lower than that of massive central LRGs at each epoch. 
Interestingly, while the parameter $\Mmin$ does not depend on redshift, $M_{1}$ exhibits a slight redshift evolution such that the LRGs at lower redshift have slightly higher $M_{1}$ for a given galaxy number density. 

\begin{figure*}[tbp]
\plotone{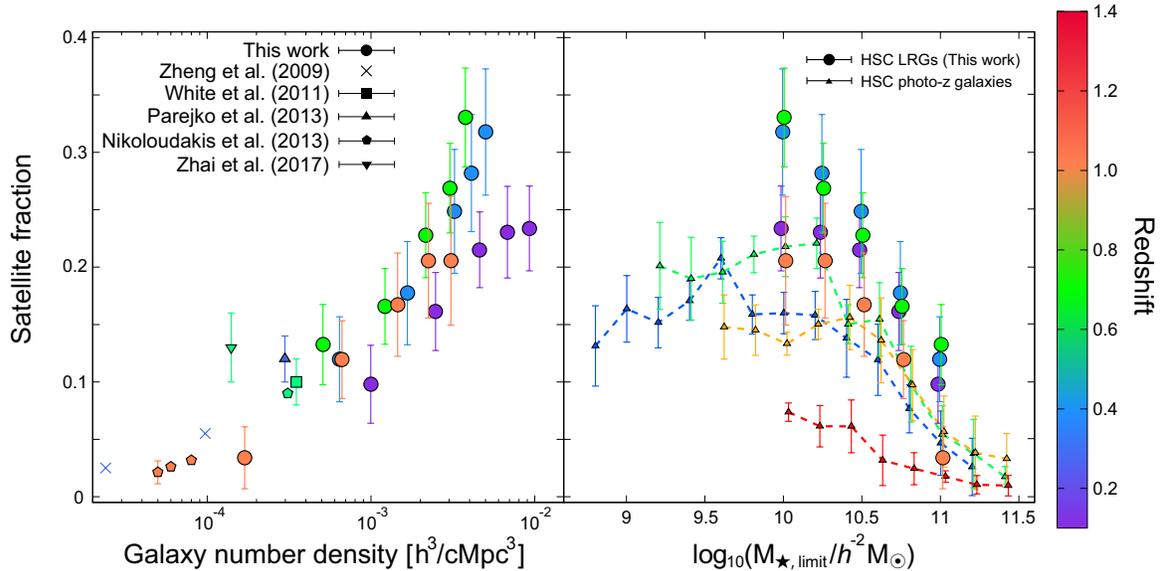}
\caption{Similar to Figure~\ref{fig:mh}, but for the satellite fraction. }
\label{fig:fsat}
\end{figure*}

Next, in the right panel of Figure~\ref{fig:mh} we compare our constraints on the HOD mass parameters for LRGs to those for the different galaxy population, the photo-$z$ galaxies, obtained from the same HSC survey \citep{ishikawa19}. 
The stellar masses and photometric redshifts of the HSC photo-$z$ galaxies are evaluated through a template spectral energy distribution-fitting technique with Bayesian priors on the galaxy physical properties \citep[{\sc Mizuki};][]{tanaka15}. 
The photo-$z$ galaxies are selected using the same HSC SSP S16A Wide layer dataset as the one used in this paper, but without imposing any color cuts. Hence, our LRG sample is a subset of the HSC photo-$z$ galaxy sample that includes both red and blue galaxies. 
Since the assumed initial mass functions (IMFs) for the stellar-mass estimation are different between the LRGs and photo-$z$ galaxies such that stellar masses of the LRGs are evaluated assuming the Salpeter IMF \citep{salpeter55} and those of photo-$z$ galaxies are estimated assuming the Chabrier IMF \citep{chabrier03}, we multiply by a factor of $1.65$ to the stellar masses of the photo-$z$ galaxies to account for the offset of stellar masses introduced by the different IMFs. 

One of the largest differences between these two populations is that $\Mmin$ of LRGs increases less rapidly with increasing the stellar-mass threshold for all the redshift bins except for the lowest-$z$ bin, indicating that the dark halo mass is the key parameter for the formation of central LRGs. 
There are many quenching models that explain the formation of red-sequence galaxies, including the mass quenching \citep[e.g.,][]{peng10,peng12,geha12} and the environmental quenching \citep[e.g.,][]{gunn72,vdbosch08,wetzel13}, and various physical mechanisms, such as a hot-halo quenching due to the virial-shock heating \citep[e.g.,][]{birnboim03,dekel09} and the radio-mode AGN feedback \citep[e.g.,][]{keres09,gabor11}, have been proposed to quench star formation according to their stellar masses and/or dark halo masses. 
We will discuss the implication of our results on the quenching models and the formation scenario of LRGs in Section~\ref{sec:discussion}. 

In addition, values of $\Mmin$ for LRGs are significantly larger than those for the photo-$z$ galaxies at the low stellar mass end. 
Since the HSC photo-$z$ galaxies consist of both star-forming and passive galaxies, this result suggests that low-mass central LRGs reside in more massive dark halos compared to the photo-$z$ galaxies with similar stellar masses, for which star-forming galaxies dominate the overall photo-$z$ galaxy population. 

In contrast to $\Mmin$, $M_{1}$ of LRGs shows the stellar-mass dependence similar to that obtained for the photo-$z$ galaxies. 
Our results show that $M_{1}$ of LRGs evolves little with redshift at a fixed stellar mass, whereas those of photo-$z$ galaxies appear to have stronger redshift dependence. 
However, since $M_{1}$ contains relatively large uncertainties compared to $\Mmin$, it is difficult to draw a definitive conclusion from the current observational results. 

\begin{figure*}[tbp]
\plotone{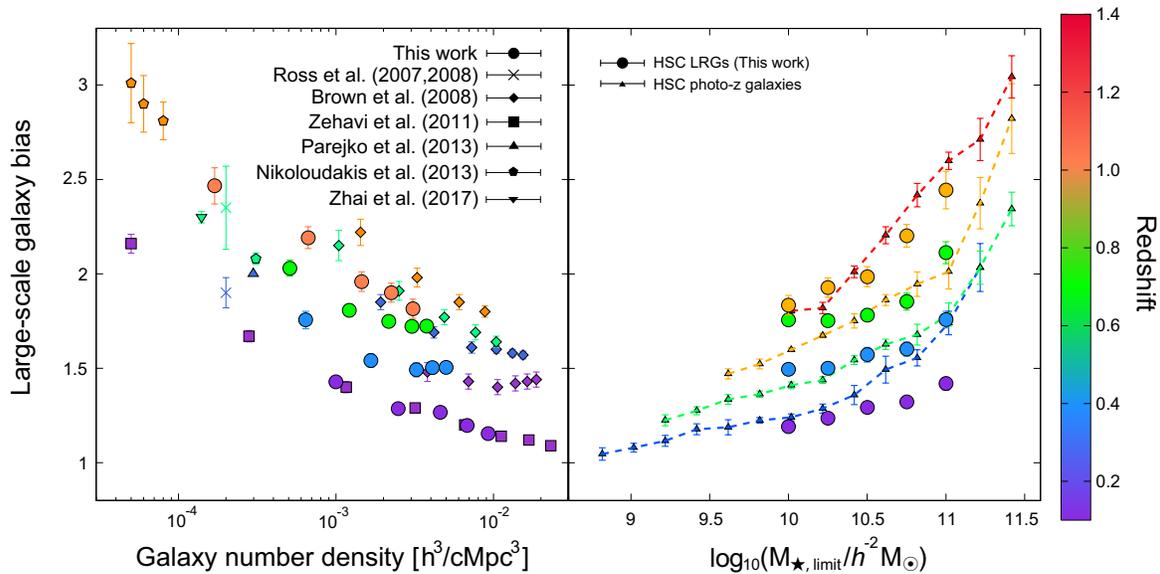}
\caption{Similar to Figure~\ref{fig:mh}, but for the large-scale galaxy bias. }
\label{fig:bg}
\end{figure*}

\subsection{Satellite fraction} \label{subsec:fsat}
In this subsection we focus on the fraction of LRGs that are satellites, which can be determined from the constrained HOD parameters. 
Specifically, the satellite fraction $\fsat$ can be calculated as:
\begin{eqnarray}
\fsat & = & 1 - f_{\rm c} \nonumber \\
 & = & 1 - \frac{1}{n_{g}^{\rm HOD}} \int d\Mh \frac{dn}{d\Mh} N_{\rm c}\left( \Mh \right), 
\label{eq:fs}
\end{eqnarray}
where $f_{\rm c}$ represents the fraction of central galaxies. 
In the left panel of Figure~\ref{fig:fsat}, we show the satellite fraction of LRGs as a function of the galaxy number density and compare them with those in the literature. 
While there are no data from previous studies in the high number density region explored in this paper ($n_{g}>10^{-3}$ $(h/{\rm cMpc})^{3}$), satellite fractions obtained by \citet{zheng09}, \citet{white11}, and \citet{nikoloudakis13} appear to be consistent with the extrapolations of our $\fsat$ values to the lower galaxy number densities. 
The satellite fraction of \citet{zhai17} is larger than those of the other studies as well as the extrapolation of our result, although the discrepancy is at most $\sim1.5\sigma$ level and is not significant. 

Differences of the satellite fraction between the LRGs and photo-$z$ galaxies are presented in the right panel of Figure~\ref{fig:fsat}. 
We find that, when compared for the same stellar mass limit, satellite fractions of the LRGs are higher than those of the photo-$z$ galaxies irrespective of their stellar-mass and redshift ranges, indicating that systems that consist of a central LRG and satellite LRGs are ubiquitous at least up to $z=1.05$. 
Using a SDSS group catalog and cosmological $N$-body simulations, \citet{wetzel13} proposed a ``delayed-then-rapid'' quenching scenario, in which SFRs of infalled satellites keep evolving for a few Gyr after the infall, and then they quench with a short time scale. 
In this scenario, stellar masses of the infalled satellites can grow as much as those of the centrals in the same halo. 
This scenario can explain our observational results that the satellite fraction of LRGs is higher than that of photo-$z$ galaxies and drastically increases with redshift. 

Observational studies have found that physical characteristics of galaxies correlate with nearby galaxies and/or their environments, known as a galactic conformity effect \citep[e.g.,][]{weinmann06,kauffmann10}. 
The $1$-halo conformity, which is an association of physical properties of central galaxies with those of satellite galaxies within the same dark halos, is well studied at $z<1$, which indicates that passive satellite galaxies are likely to be hosted by passive central galaxies \citep{hartley15,berti17}. 
The high values of LRG satellite fractions imply that the environmental quenching is effective for the wide stellar-mass range even for LRGs. 
In addition, LRGs are known as a galaxy population that formed in high-redshift Universe and observed in their old and passively evolving phase \citep[e.g.,][]{eisenstein01}. 
Therefore, the high satellite fraction of LRGs may also be explained by merging events experienced in their long evolving history. 

\subsection{Galaxy bias} \label{subsec:bg}
In this subsection we discuss the relation of the spatial clustering between galaxies and underlying dark matter. 
It is quantified by a large-scale galaxy bias, $\bg$, which can be evaluated through a set of given HOD parameters as:
\begin{equation}
\bg = \frac{1}{n_{g}^{\rm HOD}} \int d\Mh \bh\left(\Mh\right) \frac{dn}{d\Mh} N\left( \Mh \right), 
\label{eq:bgal}
\end{equation}
where $\bh \left(\Mh\right)$ is the large-scale halo bias \citep{tinker10}. 

The galaxy biases calculated for our LRG sample using the equation~(\ref{eq:bgal}) and those in the literature are presented in the left panel of Figure~\ref{fig:bg}. 
Our results are in good agreement with those of \citet{ross07,ross08}, \citet{zehavi11}, \citet{parejko13}, \citet{nikoloudakis13}, and \citet{zhai17}. 
However, galaxy biases measured by \citet{brown08} are systematically larger than those in the other studies. 
A possible reason of this discrepancy is the difference of procedures to select LRGs. 
Studies using data obtained in the SDSS adopt a color selection described in \citet{eisenstein01}, which is based on the SDSS $ugriz$ optical magnitudes \citep{fukugita96}, and the CAMIRA algorithm is also designed to select red-sequence galaxies with colors similar to SDSS LRGs. 
However, \citet{brown08} selected red galaxies from the bimodal galaxy distribution in the rest-frame $(U-V)$ color versus $V$-band absolute magnitude diagram presented by \citet{bell04}. 
Therefore, physical characteristics of galaxies of \citet{brown08} can be slightly different from those in other LRG studies, which might explain the difference mentioned above. 
In addition, the difference of the cosmological parameters may partly explains the discrepancy. 
To check this possibility, we repeat our HOD analysis of the LRG sample at each redshift bin adopting the {\it WMAP3} cosmologies \citep{spergel07} as adopted in \citet{brown08}, and find that the resulting galaxy biases increase by a few percents on average compared to our original analysis, and hence reduces the discrepancy between our result and the \citet{brown08} result. 

We compare the galaxy biases of the LRGs with those of the photo-$z$ galaxies in the right panel of Figure~\ref{fig:bg}. 
Our results indicate that galaxy biases of low-mass LRGs satisfying $\Mstar \lesssim 10^{10.5}h^{-2}\Msun$ depend only weakly on stellar masses and the galaxy biases rapidly increases with increasing stellar masses at the high stellar mass end. 
While the stellar-mass dependence of the photo-$z$ galaxies shows a trend similar to the LRGs, galaxy biases of the photo-$z$ galaxies at $10^{10.0} \lesssim \Mstarlimit/h^{-2}\Msun \lesssim 10^{10.5}$ show stronger dependence on the stellar mass, which is consistent with the trend found for $\Mmin$. 
The rapid increase of the galaxy biases of the LRGs indicates that only the most massive LRGs are highly biased objects in low-$z$, whereas even intermediate-mass LRGs $(\Mstar \sim 10^{10.5}h^{-2}\Msun)$ are rare at $z \sim 1$. 

\section{Discussion} \label{sec:discussion}
\subsection{Correlation between the LRG formation and the dark halo mass}
As shown in Figure~\ref{fig:fsat}, the satellite fraction of LRGs is much higher compared to the all photo-$z$ sample when compared for the same stellar masses, indicating that LRGs tend to reside in more dense regions. 
Observational studies have found that the galaxy quenching is much more efficient in dense regions \citep[e.g.,][]{peng10,peng12}, and this environmental quenching effect can be triggered by the galaxy harassment \citep{moore96} and the ram-pressure stripping \citep{gunn72}. 
Therefore, the environmental quenching that is efficient in high-density environments can play an important role in the formation of LRGs. 

Interestingly, we also find that dark halo masses of central LRGs (i.e., $\Mmin$) is tightly correlated with the number density of LRGs in all the redshift range we examined, $0.1<z<1.05$ (Figure~\ref{fig:mh}), which may not be explained well by the environmental quenching that is caused by the interactions with surrounding galaxies. 
This may imply that the global environment as well as the local environment can have an impact on the galaxy quenching. 

In addition, Figure~\ref{fig:mh} shows that halo masses of central LRGs are almost constant at $\Mstar \lesssim 10^{10.75}h^{-2}\Msun$ at $z>0.3$, which supports the idea that the halo mass is the key parameter for the galaxy quenching mechanism and the formation of LRGs. 
The halo quenching is also the galaxy quenching model caused by the disturbance of star-forming activities due to e.g., the virial shock heating \citep[e.g.,][]{birnboim03,dekel06}. 
Our results suggests that $\Mh \sim 10^{12.5 \pm 0.2}h^{-1}\Msun$ is the threshold dark halo mass such that galaxies hosted by those dark halos are efficiently transformed into the LRGs irrespective of the baryonic properties, which is consistent with the conclusion of \citet{woo13}. 

In short, the formation of LRGs is largely connected to the environment of host dark halos of LRGs, but the internal effect, such as the AGN feedback, can also contribute to the suppression of star formation to some extent. 
Our results suggest that $\Mh \sim 10^{12.5 \pm 0.2}h^{-1}\Msun$ is the critical halo mass for the formation of LRGs, at least for LRGs with stellar masses of $\Mstar \lesssim 10^{10.75}h^{-2}\Msun$ at $z>0.3$. 
We leave the exploration of the different behavior of the stellar mass dependence of halo masses at $z<0.3$ to future work. 

\subsection{Mass assembly history and evolution of LRGs}
The evolution of the dark halo mass also provides a clue to the evolution of galaxies. 
In this subsection, we trace the evolution of median dark halo masses of central LRGs, i.e., $\Mmin$, from $z\sim1$ to $z=0$ for connecting our LRGs at the highest-$z$ bin with those at lower-$z$ bins using the IllustrisTNG simulation \citep{springel18}. 

We use the TNG100 simulation, which well reproduces observational stellar-mass functions even at $z\sim1$ \citep{pillepich18}. 
The definition of the stellar mass is the total mass of stellar particles within twice the stellar half mass radius obtained by the \textsc{subfind} algorithm \citep{springel01}, while that of the halo mass is the total mass of dark matter particles that consists of dark halos identified by the friend-of-friend algorithm. 
Red-sequence galaxies from the TNG100 simulation (hereafter TNG LRGs) are selected according to their stellar masses and star-formation rates, which is in the same manner as in \citet{moustakas13}, and we select them only at $z=0.92$ that corresponds to the effective redshift of our $z_{4}$ redshift bin. 
Three stellar-mass limited samples satisfying $\logMstarlimit \geq$ 10.0, 10.5, and 11.0 are prepared. For each sample, we trace the halo mass of each TNG LRGs from $z=0.92$ to $0$ and derive the evolutionary history of the median halo mass of each stellar-mass limited sample. 

\begin{figure}[tbp]
\epsscale{1.0}
\plotone{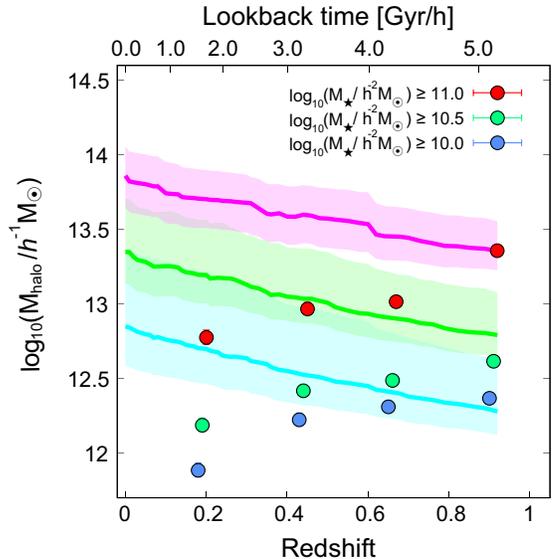}
\caption{Evolution of median dark halo masses as a function of redshift. Circles show $\Mmin$ of the HSC LRGs with different stellar-mass thresholds. Lines indicate median halo mass accretion histories traced by the IllustrisTNG simulation and shade regions show the root mean squares of the halo masses in the simulation at each redshift. Specifically, LRG-like galaxies in the IllustrisTNG are selected by imposing the same stellar-mass threshold of circles with corresponding colors at $z=0.92$, and the evolution of halo masses of host halos of these LRGs are traced from $z=0.92$ to $0$. Our result implies that a halo which hosts LRGs with the stellar mass of $\Mstar\geq 10^{10}h^{-2}\Msun$ at $z=0.9$ tends to host those with $\Mstar\geq 10^{11}h^{-2}\Msun$ at $z=0$. }
\label{fig:mh_accretion}
\end{figure}

Figure~\ref{fig:mh_accretion} shows the evolution of the median halo masses and the root mean square of masses of the TNG LRGs. 
Again, we emphasize that in the TNG LRGs are selected only at $z=0.92$ satisfying the stellar-mass limits of $\logMstarlimit \geq$ $10.0$ (blue), $10.5$ (green), and $11.0$ (red), and evolutionary tracks of halo masses indicate the assembly history of halo masses those LRGs from $z=0.92$ to today. 

First, we check the consistency of the median halo masses at $z=0.92$. 
We find that the median halo masses of our observed LRGs show excellent agreement with those from the simulation, which indicates the consistency between the observation and the simulation. 
Then we investigate the relation between HSC LRGs at $z=0.92$ and $z<0.92$ by comparing the observed halo masses with the evolutionary history of the median halo masses from $z=0.92$ to $0$. 
By matching halo masses at each redshift bin, we find that the low-mass LRGs at $z_{4}$ ($z\sim 0.92$) evolve into the intermediate-mass LRGs at $z_{3}$ ($z\sim 0.65$) and $z_{2}$ ($z\sim 0.45$) bins and finally evolve into the high-mass LRGs at $z_{1}$ ($z\sim 0.2$). 
While LRGs are thought to be passively evolving galaxies, the halo mass accretion history indicates that low-mass LRGs at $z=0.92$ increase their stellar masses by an order of magnitude from $z=0.92$ to $0$. 
Next, we check the evolution of high-mass LRGs at $z=0.92$. 
The halo mass of high-mass central LRGs at $z_{4}$ is expected to reach $\sim 10^{14}h^{-1}\Msun$ at $z=0$, which corresponds to the halo masses of galaxy clusters. 
Therefore, progenitors of BCGs of local clusters can be the massive LRGs at $z\sim1$. 

However, previous studies have shown that LRGs increase their stellar masses at most $\sim 30-50$\% through mergers since $z\sim1$ \citep{cool08,skelton12,lopez-sanjuan12}. In addition, \citet{banerji10} found that massive LRGs with stellar masses of $\Mstar>10^{11}\Msun$ complete their stellar mass assembly by $z\sim0.8$. 
These observational results appear to be inconsistent with the evolutionary scenario of the low-mass LRGs that we speculated above via the halo mass evolution. 
This discrepancy can be explained by a large contribution of LRGs from other galaxy populations. 
Put another way, our result implies that most of massive ($\logMstarlimit \geq 11$) LRGs at $z\sim 0.2$ are not descendants of LRGs with similar stellar masses at $z\sim1$, but they are recently migrated from other galaxy populations. 
The large difference of their number densities shown in Table~\ref{tab:nz} supports this interpretation. 
One possible scenario is the contribution from green valley galaxies (or star-forming galaxies), which are not included in our LRG sample but can turn into LRGs by an efficient quenching. 

As mentioned above, the majority of the low-mass LRGs is not expected to experience significant stellar mass grow from $z\sim 1$ to $0$. 
However it is still possible that some minor fraction of massive LRGs evolved from less massive LRGs through mergers and satellite. 
The redshift dependence of the satellite fraction (the right panel of Figure~\ref{fig:fsat}) shows an increasing trend from $z\sim1$ to $z\sim0.3$ and then decreases to $z\sim 0.1$ at a fixed stellar-mass threshold. 
Therefore possible scenario that partly explains our results is that low-mass LRGs accrete onto more massive halos to become satellite galaxies and also grow their stellar masses by mergers at $0.3<z<0.9$, and then those satellite LRGs merge into central LRGs at $z<0.3$. 

\section{Summary} \label{sec:summary}
We have presented the clustering analysis of LRGs selected in the HSC SSP S16A Wide layer covering $\sim 124$ deg$^{2}$ \citep{aihara18}. 
We have used $615,317$ LRGs selected by the CAMIRA algorithm \citep{oguri14} with photometric redshift and stellar mass measurements to investigate the dependence of clustering and physical properties of LRGs on their redshifts and the stellar masses. 
We have derived, for the first time, the relation between LRGs and their host dark halos as a function of their stellar masses using the HOD formalism and compared them with those obtained in the literature to check the consistency. 
Our results have also been compared with those of the photo-$z$ galaxies including both red and blue galaxies selected from the same HSC SSP S16A Wide layer \citep{ishikawa19} to highlight the difference between LRGs and normal galaxies. 

The major findings of this study through the HOD analysis are summarized as follows: 
\begin{enumerate}
\item 
The characteristic dark halo masses of central galaxies $\Mmin$ is tightly correlated with the number densities of the LRGs irrespective of redshifts, which is consistent with previous studies. 
The mass of a halo to host one LRG satellite galaxy, $M_{1}$, also follows a power-law relation as a function of the number density as in the case of $\Mmin$, although the relation is not as tight as for $\Mmin$ and shows some redshift evolution. 

\item
The characteristic dark halo mass $\Mmin$ depends only weakly on the stellar mass at $\Mstar \lesssim 10^{10.75}h^{-2}\Msun$, which indicates that the dark halo mass is the key parameter for the formation of the LRGs rather than the stellar mass. 
We have found that $\Mh \sim 10^{12.5 \pm 0.2} h^{-1}\Msun$ is the critical dark halo mass for the LRG formation, at least for LRGs with stellar masses $\Mstar \lesssim 10^{10.75}h^{-2}\Msun$ at $0.3<z<1.05$. 
This threshold halo mass is expected to be originated from the halo quenching mechanism due to the halo environment. 

\item 
The satellite fractions of the LRGs are much higher compared to those of the photo-$z$ galaxy sample all the redshift and stellar mass ranges examined in this paper, which indicates that the LRGs tend to reside in high density environments even at $z \sim 1$. 
Moreover, the high satellite fractions of LRGs are indicative of the $1$-halo galactic conformity up to $z\sim1$. 

\item 
We have found that the large-scale galaxy bias of LRGs monotonically increases with increasing stellar mass and redshift. 
At a fixed redshift, the increasing trend drastically changes at the massive end such that only the most massive LRGs are highly biased objects in low-$z$ and even the intermediate-mass LRGs $(\Mstar \sim 10^{10.5}h^{-2}\Msun)$ are rare at $z \sim 1$. 

\item
We compare the observed median halo masses of central HSC LRGs $(\Mmin)$ with those derived from the IllustrisTNG simulation. 
The median halo masses of our HSC LRGs at $0.8\leq z \leq 1.05$ calculated by the HOD analysis show excellent agreement with those in the simulation at $z=0.92$. 
By comparing our results with the evolution of halo masses from $z=0.92$ to $0$ in the simulation, we argue that low-mass LRGs at $z\sim1$ can evolve into intermediate-mass LRGs at $0.3\lesssim z \lesssim0.8$ and high-mass LRGs at $z\sim0.2$. Such stellar mass growth may be realized by galaxy mergers and satellite accretions. On the other hand, the comparison suggests that massive LRGs at low redshifts are mainly formed from green valley galaxies or evolved from less massive LRGs through mergers and satellite accretions. 

\end{enumerate}

This paper has presented the first clustering analysis of the HSC LRGs and demonstrated its power for studying the large-scale structure. One possible application of the HSC LRGs may be the detection of baryon acoustic oscillations \citep[BAO;][]{eisenstein05,okumura08} to constrain cosmological parameters. While the sample used in this paper is not large enough for this purpose, the latest LRG catalog (Oguri et al.~in prep.) covers a survey volume large enough for the detection of BAO. 
For such cosmological analysis, three-dimensional clustering in redshift space rather than the angular clustering analyzed in this paper will be more suited, as done for a similar sample by \citet{chiu20}. 
Analyzing the BAO encoded in the LRGs in the HSC survey will enable us to constrain the acoustic scale up to high redshifts, $z\lesssim 1.4$.

In addition, correlation functions in redshift space also enable us to test the theory of the structure formation based on the general relativity via redshift-space distortions \citep[e.g.,][]{davis83,guzzo08,okumura16}. 
Large and deep samples of the HSC LRG can constrain the growth rate of the Universe up to $z\sim1.4$ with high precision by analyzing the redshift-space distortion. 
Moreover, combining the two-point correlation function derived in this paper with weak-lensing signals can constrain cosmological parameters since the degeneracy between cosmological parameters and the galaxy bias can be resolved \citep[e.g.,][]{cacciato09,abbott18}. 
This paper presents a first step toward testing the structure formation theory as well as the $\Lambda$CDM cosmological model using the two-point (and higher order) statistics of unique HSC photometric LRG samples. 

\acknowledgments
We are grateful to the anonymous referee for his/her careful reading and useful comments. 
TO acknowledges support from the Ministry of Science and Technology of Taiwan under Grants No. MOST 109-2112-M-001-027- and the Career Development Award, Academia Sinica (AS-CDA-108-M02) for the period of 2019-2023. 

This paper is based on data collected at the Subaru Telescope and retrieved from the HSC data archive system, which is operated by the Subaru Telescope and Astronomy Data Center (ADC) at National Astronomical Observatory of Japan. 
Data analysis was in part carried out with the cooperation of Center for Computational Astrophysics (CfCA), National Astronomical Observatory of Japan. 
The Subaru Telescope is honored and grateful for the opportunity of observing the Universe from Maunakea, which has the cultural, historical and natural significance in Hawaii. 

The Hyper Suprime-Cam (HSC) collaboration includes the astronomical communities of Japan and Taiwan, and Princeton University. 
The HSC instrumentation and software were developed by the National Astronomical Observatory of Japan (NAOJ), the Kavli Institute for the Physics and Mathematics of the Universe (Kavli IPMU), the University of Tokyo, the High Energy Accelerator Research Organization (KEK), the Academia Sinica Institute for Astronomy and Astrophysics in Taiwan (ASIAA), and Princeton University. 
Funding was contributed by the FIRST program from the Japanese Cabinet Office, the Ministry of Education, Culture, Sports, Science and Technology (MEXT), the Japan Society for the Promotion of Science (JSPS), Japan Science and Technology Agency (JST), the Toray Science Foundation, NAOJ, Kavli IPMU, KEK, ASIAA, and Princeton University. 

This paper makes use of software developed for the Large Synoptic Survey Telescope. 
We thank the LSST Project for making their code available as free software at  http://dm.lsst.org

The Pan-STARRS1 Surveys (PS1) and the PS1 public science archive have been made possible through contributions by the Institute for Astronomy, the University of Hawaii, the Pan-STARRS Project Office, the Max Planck Society and its participating institutes, the Max Planck Institute for Astronomy, Heidelberg, and the Max Planck Institute for Extraterrestrial Physics, Garching, The Johns Hopkins University, Durham University, the University of Edinburgh, the Queen's University Belfast, the Harvard-Smithsonian Center for Astrophysics, the Las Cumbres Observatory Global Telescope Network Incorporated, the National Central University of Taiwan, the Space Telescope Science Institute, the National Aeronautics and Space Administration under grant No. NNX08AR22G issued through the Planetary Science Division of the NASA Science Mission Directorate, the National Science Foundation grant No. AST-1238877, the University of Maryland, Eotvos Lorand University (ELTE), the Los Alamos National Laboratory, and the Gordon and Betty Moore Foundation. 

Numerical computations are in part carried out on the Cray XC50 (Aterui II) operated by the Center for Computational Astrophysics, CfCA, National Astronomical Observatory of Japan. 
Data analyses are in part carried out on the open use data analysis computer system at the Astronomy Data Center, ADC, of the National Astronomical Observatory of Japan. 

\facility{Subaru Telescope (Hyper Suprime-Cam)}.

\begin{longrotatetable}
\begin{deluxetable*}{llcccccccc}
\tablecaption{Best-fitting HOD Parameters of Cumulative Stellar-mass Limited LRG Samples}
\tablehead{Redshift & Stellar-mass limit & $\log_{10}\Mmin$ & $\log_{10}M_{1}$ & $\log_{10}M_{0}$ & $\sigmaM$ & $\alpha$ & $\fsat$ & $\bg$ & $\chi^{2}$/d.o.f.}
\startdata
$0.10 \leq z < 0.30$ & $10.0$ & $11.884^{+0.049}_{-0.044}$ & $13.069^{+0.111}_{-0.076}$ & $10.612^{+1.326}_{-2.521}$ & $0.490^{+0.247}_{-0.310}$ & $1.053^{+0.125}_{-0.290}$ & $0.234 \pm 0.037$ & $1.154 \pm 0.019$ & $0.58$ \\
 & $10.25$ & $12.020^{+0.048}_{-0.041}$ & $13.179^{+0.116}_{-0.078}$ & $10.784^{+1.306}_{-3.120}$ & $0.479^{+0.283}_{-0.331}$ & $1.061^{+0.172}_{-0.514}$ & $0.230 \pm 0.040$ & $1.197 \pm 0.024$ & $0.61$ \\
 & $10.5$ & $12.186^{+0.049}_{-0.043}$ & $13.344^{+0.086}_{-0.063}$ & $9.072^{+2.635}_{-2.833}$ & $0.487^{+0.255}_{-0.294}$ & $1.138^{+0.120}_{-0.319}$ & $0.215 \pm 0.033$ & $1.266 \pm 0.021$ & $0.19$ \\
 & $10.75$ & $12.400^{+0.051}_{-0.042}$ & $13.651^{+0.123}_{-0.113}$ & $9.261^{+2.772}_{-3.025}$ & $0.453^{+0.266}_{-0.277}$ & $1.078^{+0.188}_{-0.386}$ & $0.161 \pm 0.034$ & $1.285 \pm 0.020$ & $0.13$ \\
 & $11.0$ & $12.774^{+0.051}_{-0.045}$ & $14.093^{+0.312}_{-0.129}$ & $9.053^{+2.836}_{-2.769}$ & $0.476^{+0.262}_{-0.289}$ & $1.207^{+0.343}_{-0.401}$ & $0.098 \pm 0.034$ & $1.428 \pm 0.027$ & $0.15$ \\
\hline
$0.30 \leq z < 0.55$ & $10.0$ & $12.224^{+0.044}_{-0.044}$ & $13.064^{+0.083}_{-0.065}$ & $8.602^{+2.360}_{-2.503}$ & $0.545^{+0.308}_{-0.356}$ & $1.119^{+0.102}_{-0.115}$ & $0.318 \pm 0.055$ & $1.505 \pm 0.030$ & $0.10$ \\
 & $10.25$ & $12.290^{+0.041}_{-0.040}$ & $13.178^{+0.077}_{-0.063}$ & $8.745^{+2.454}_{-2.584}$ & $0.550^{+0.299}_{-0.352}$ & $1.137^{+0.076}_{-0.163}$ & $0.282 \pm 0.051$ & $1.502 \pm 0.030$ & $0.11$ \\
 & $10.5$ & $12.418^{+0.042}_{-0.040}$ & $13.313^{+0.098}_{-0.086}$ & $10.244^{+2.038}_{-4.372}$ & $0.581^{+0.302}_{-0.410}$ & $1.090^{+0.145}_{-0.578}$ & $0.249 \pm 0.054$ & $1.492 \pm 0.037$ & $0.49$ \\
 & $10.75$ & $12.607^{+0.042}_{-0.041}$ & $13.637^{+0.125}_{-0.104}$ & $9.489^{+2.795}_{-3.175}$ & $0.543^{+0.318}_{-0.340}$ & $1.125^{+0.212}_{-0.444}$ & $0.177 \pm 0.045$ & $1.541 \pm 0.036$ & $0.21$ \\
 & $11.0$ & $12.967^{+0.044}_{-0.039}$ & $14.020^{+0.113}_{-0.088}$ & $9.423^{+2.983}_{-3.078}$ & $0.540^{+0.317}_{-0.341}$ & $1.310^{+0.266}_{-0.447}$ & $0.120 \pm 0.037$ & $1.755 \pm 0.046$ & $0.27$ \\
\hline
$0.55 \leq z < 0.80$ & $10.0$ & $12.311^{+0.039}_{-0.036}$ & $13.070^{+0.067}_{-0.055}$ & $8.480^{+2.293}_{-2.431}$ & $0.494^{+0.240}_{-0.262}$ & $1.113^{+0.065}_{-0.075}$ & $0.330 \pm 0.043$ & $1.723 \pm 0.026$ & $0.08$ \\
 & $10.25$ & $12.366^{+0.041}_{-0.036}$ & $13.227^{+0.058}_{-0.048}$ & $8.592^{+2.466}_{-2.460}$ & $0.497^{+0.241}_{-0.269}$ & $1.182^{+0.081}_{-0.090}$ & $0.269 \pm 0.039$ & $1.722 \pm 0.028$ & $0.08$ \\
 & $10.5$ & $12.486^{+0.042}_{-0.037}$ & $13.392^{+0.078}_{-0.059}$ & $8.834^{+2.559}_{-2.645}$ & $0.497^{+0.236}_{-0.268}$ & $1.186^{+0.092}_{-0.224}$ & $0.228 \pm 0.037$ & $1.749 \pm 0.029$ & $0.10$ \\
 & $10.75$ & $12.671^{+0.039}_{-0.034}$ & $13.673^{+0.077}_{-0.074}$ & $9.113^{+2.691}_{-2.808}$ & $0.486^{+0.253}_{-0.258}$ & $1.180^{+0.119}_{-0.311}$ & $0.166 \pm 0.033$ & $1.807 \pm 0.033$ & $0.21$ \\
 & $11.0$ & $13.016^{+0.039}_{-0.037}$ & $13.991^{+0.094}_{-0.094}$ & $9.517^{+2.953}_{-3.155}$ & $0.520^{+0.215}_{-0.280}$ & $1.176^{+0.198}_{-0.381}$ & $0.133 \pm 0.035$ & $2.030 \pm 0.042$ & $0.17$ \\
\hline
$0.80 \leq z \leq 1.05 $ & $10.0$ & $12.367^{+0.038}_{-0.035}$ & $13.202^{+0.053}_{-0.043}$ & $8.410^{+2.306}_{-2.250}$ & $0.636^{+0.245}_{-0.499}$ & $1.265^{+0.076}_{-0.074}$ & $0.205 \pm 0.056$ & $1.815 \pm 0.051$ & $0.54$ \\
 & $10.25$ & $12.450^{+0.037}_{-0.034}$ & $13.297^{+0.057}_{-0.048}$ & $8.368^{+2.304}_{-2.316}$ & $0.575^{+0.298}_{-0.402}$ & $1.257^{+0.078}_{-0.078}$ & $0.206 \pm 0.050$ & $1.900 \pm 0.050$ & $0.46$ \\
 & $10.5$ & $12.614^{+0.039}_{-0.034}$ & $13.486^{+0.075}_{-0.058}$ & $8.910^{+2.557}_{-2.652}$ & $0.589^{+0.279}_{-0.347}$ & $1.265^{+0.088}_{-0.502}$ & $0.167 \pm 0.045$ & $1.959 \pm 0.052$ & $0.22$ \\
 & $10.75$ & $12.818^{+0.041}_{-0.034}$ & $13.788^{+0.080}_{-0.066}$ & $9.076^{+2.743}_{-2.711}$ & $0.499^{+0.322}_{-0.299}$ & $1.369^{+0.212}_{-0.292}$ & $0.119 \pm 0.034$ & $2.192 \pm 0.058$ & $0.08$ \\
 & $11.0$ & $13.357^{+0.038}_{-0.035}$ & $14.479^{+0.317}_{-0.197}$ & $9.428^{+3.096}_{-3.021}$ & $0.635^{+0.245}_{-0.399}$ & $1.346^{+0.391}_{-0.407}$ & $0.034 \pm 0.027$ & $2.466 \pm 0.096$ & $0.12$ \\
\enddata
\tablecomments{The stellar-mass limit is in units of $h^{-2}\Msun$ in a logarithmic scale, whereas all halo-mass parameters are in units of $h^{-1}\Msun$.}
\label{tab:hod_params}
\end{deluxetable*}
\end{longrotatetable}

\end{document}